\documentclass[reprint, aps, longbibliography]{revtex4-1}
\usepackage{amsmath}
\usepackage{verbatim}
\usepackage{amssymb}
\usepackage{setspace}
\usepackage{graphicx}
\usepackage{geometry}
\usepackage{bm}
\usepackage{float}
\usepackage{color,soul}
\def\be{\begin{equation}}
\def\ee{\end{equation}}
\def\bea{\begin{eqnarray}}
\def\eea{\end{eqnarray}}

\begin{document}
\preprint{APS/123-QED}
\title{Effects of Cosmic String Velocities and the Origin of Globular Clusters}
\author{Ling Lin}
   \email{ling.lin2@mail.mcgill.ca}
\author{Shoma Yamanouchi}
   \email{shoma.yamanouchi@mail.mcgill.ca}
\author{Robert Brandenberger}
   \email{rhb@physics.mcgill.ca}
\affiliation{Department of Physics, McGill University, Montr$\acute{e}$al, QC, H3A 2TS, 
Canada}

\begin{abstract}

With the hypothesis that cosmic string loops act as seeds for globular clusters in mind, 
we study the role that velocities of these strings will play in determining the 
mass distribution of globular clusters. Loops with high enough velocities will 
not form compact and roughly spherical objects and can hence not be the seeds for
globular clusters. We compute the expected number density 
and mass function of globular clusters as a function of both the string tension and
the peak loop velocity, and compare the results with the observational 
data on the mass distribution of globular clusters in our Milky Way. We 
determine the critical peak string loop velocity above which the agreement
between the string loop model for the origin of globular clusters (neglecting
loop velocities) and observational data is lost.

\end{abstract}

\maketitle

\section{\label{sec:1}Introduction}

We have recently \cite{us} made the hypothesis that string loops arising
from a scaling network of cosmic strings seed the formation of the globular clusters 
which are observed to be distributed in the halos of galaxies, in particular our
own Milky Way galaxy. Our model easily explains the observational facts that
globular clusters are the oldest and most compact star clusters in our galaxy,
and that they are distributed throughout the halo as opposed to only in the
disk. 

Cosmic string loops arise via the interaction of infinite string segments
which in turn are generated during a symmetry breaking phase transition in
the early universe. Cosmic strings are predicted in a large class of models
of particle physics beyond the {\it Standard Model}. According to the
cosmic string scaling solution \cite{Kibble}, the distribution of infinite
string segments is independent of the cosmic string tension $\mu$ (which is,
in the natural units we use, equal to the mass per unit length). As a 
consequence, cosmic string loops are also formed with a number density
which is independent of $\mu$. The distribution of string loops is
determined by the string scaling solution, and depends on $\mu$ only
through the dependence on $\mu$ of a critical loop radius $R_c$ below
which the number distribution of strings becomes constant \cite{CSearly}.

In our previous paper \cite{us} we have shown that if we fix the one free parameter
in our theoretical model, namely the string tension $\mu$, then
the peak number density and the mass distribution are fixed. Demanding
that the mass distribution peaks at a value corresponding to the
peak in the observed mass function of globular clusters in the Milky
Way gave us a value of $G \mu \sim 10^{-9.5}$ (where we - as is standard
in the cosmic string literature - multiplied $\mu$
by Newton's gravitational constant $G$ in order to obtain a dimensionless
number), a value which is below the current upper bound on $G \mu$ of
$G \mu < 1.5 \times 10^{-7}$ \cite{CSlimit} (see also 
\cite{others} for earlier work on limits on the string tension). At this point, our
string model had no more free parameters. Interestingly, we found good
agreement between the predicted and observed globular cluster mass functions. 

However, in our previous study \cite{us} we neglected the presence
of string loop velocities (center of mass string loop velocities).
Recent numerical simulations (eg. \cite{VSimulations}) tell us that 
loops are typically born with translational velocities that are sizable 
fractions of the speed of light. The reason is that, since long 
string segments usually have relativistic speeds, then as string loops 
split off, the loops also gain significant velocities. When velocities 
are taken into considerations, accretion  will not be spherically symmetric. 
Additionally, accretion onto a moving loop may be less efficient compared 
to accretion onto a stationary loop. It should be noted, however, that 
loop velocities also undergo red-shifting, and thus slow down as the loops age.

In the following, we first review the hypothesis that globular clusters may 
be seeded by the cosmic string loops that arise from a string scaling
solution in particle physics models with a vacuum manifold which has
the topology of a circle \cite{us}. In section \ref{sec:3} we present our 
a first analysis of velocity effects on our model for the mass function of globular 
clusters. We compute a suppression factor which takes into account that loops
with too large initial velocities will move a distance greater than the
size of the spherical object which a stationary loop would accrete, and
we incorporate this factor into the predicted overall number density to 
determine a new mass function. This is then compared with the observed mass 
distribution of globular clusters in our Milky Way galaxy. Moving
loops in fact do accrete matter, but do not give rise to a spherical
distribution. In Section \ref{sec:4} we compare the total mass from spherical 
and non-spherical accretion, and determine an alternative criterium 
for the maximal velocity of a loop that will seed a globular cluster. The
resulting mass function turns out to be similar to the one derived
using the first criterium. We then turn to a brief discussion of the
cosmic string {\it rocket effect}, another effect neglected in our
previous work. We show that this effect has a negligible impact on
our globular cluster study. Finally, we 
present or conclusions in section \ref{sec:6}.

\section{\label{sec:2}Globular Clusters From Cosmic String Loops}

\begin{figure*}
\centering
\includegraphics[width=\textwidth]{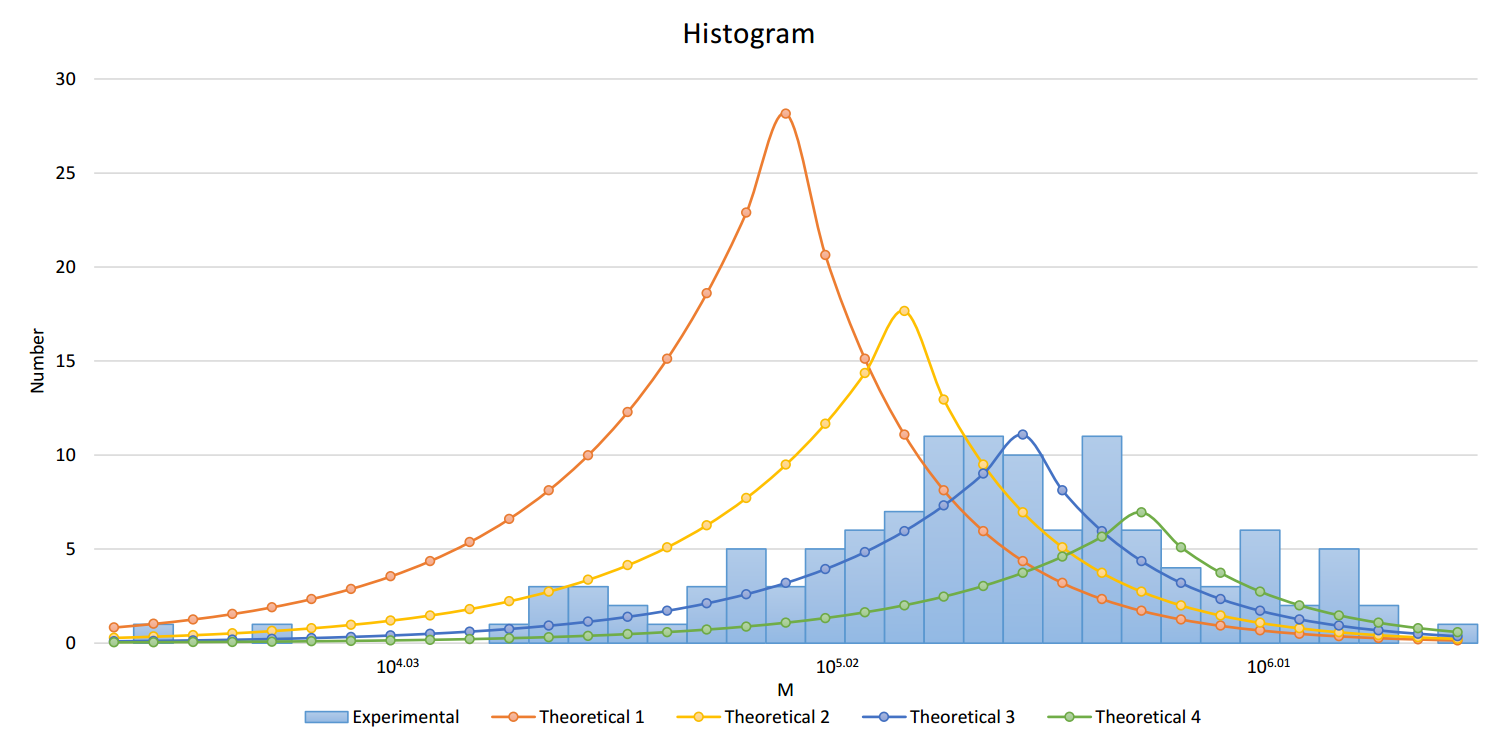}
\caption{Dependence of the mass function of our model on the string tension 
$G\mu$. The horizontal axis is mass on a logarithmic scale, the vertical axis 
gives the number density on a linear scale. The histogram show the data taken from \cite{GlobularClusterParameters}. The curves shown are for $G\mu=2.92\times10^{-10}$,
$G\mu=3.98\times10^{-10}$, $G\mu=5.43\times10^{-10}$ and $G\mu=7.41\times10^{-10}$ 
(in increasing order of mass at the peak position). The blue solid curve minimizes 
$\chi^{2}$. The cosmic string parameters chosen are described in the text. The
effects of string velocities are neglected.}
\label{fig:histogram1_2}
\end{figure*}

In this report, we will consider a one-scale model for the distribution of 
strings which \cite{Onescale1, Onescale2} implies that loops of initial radius 
$R_{i}$ form at  an initial time $t_{i}$ given by
\begin{equation}
   \label{eqn:ti}
   t_{i} \, = \, \alpha^{-1}\beta R_{i} \, .
\end{equation}
Here, $\alpha$ is a constant obtained from numerical solutions \cite{VSimulations}.
We will use the value $\alpha = 0.3$. The average length of a string loop is 
given by $l=\beta R$, where $\beta$ is taken to be $10$ here; if the string 
loops were exactly circular $\beta$ would be $2\pi$. Making use of the 
fast decay approximation that assumes loops decay virtually instantaneously, 
we have $R(t)=R_{i}(t_{i})$ until the decay time. String loops oscillate
and decay by the emission of gravitational radiation \cite{grav} whose strength is
parametrized by a dimensionless constant $\gamma$. Loops with radius smaller
than 
\be
R_c(t) \, = \, \gamma G \mu t 
\ee
live less than one Hubble expansion time before decaying. Hence, the
number density of loops is constant for $R < R_c(t)$.
 
Linear cosmological perturbation theory tells us that accretion of matter 
around a cosmic string loop starts at $t_{eq}$, the time of equal
matter and radiation. At this time, the mass 
function of the string loop with respect to radius is dominated by loops 
with a critical radius $R_{c_{1}}$, given by
\begin{equation}
   \label{eqn:Rc1}
   R_{c_{1}} \, = \, \gamma G\mu t_{eq} \, .
\end{equation}
We will use $\gamma=10^{2}$. Though these cosmic string loops will have 
decayed by the present time, the objects they seed will continue to grow. 
The number density at a time $t > t_{eq}$ for loops formed in the 
radiation dominated era of 
such objects inside a galaxy is given by (see e.g. \cite{CSrevs} for
reviews on the applications of cosmic strings in cosmology)
\begin{align}
   \label{eqn:numberdensity}
   n(R,t)&=N\alpha^{5/2}\beta^{-5/2}t_{eq}^{1/2}t^{-2}R^{-5/2}\quad\text{for}\;R>R_{c_{1}} \nonumber \\
   n(R,t)&=N\alpha^{5/2}\beta^{-5/2}\gamma^{-5/2}(G\mu)^{-5/2}t_{eq}^{-2}t^{-2} \nonumber \\
   &=\text{const.}\quad\text{for}\;R<R_{c_{1}}
\end{align}
The constant $N$ depends on the square of the average number $\tilde{N}$ 
of long string segments per Hubble volume since two long string segments are 
required to form a loop. We will take $N=10^{2}$. Incorporating the 
Zel’dovich approximation \cite{Zeldovich}, the local number density of 
string loops inside a galaxy will be enhanced by a factor of $F$. This 
factor is estimated to be $F=64$ due to accretion and virialization in each 
direction. In our calculations, we will use $F=10^{2}$.

Assuming accretion continues to the present time, the mass which has accreted 
about these seed loops at the present time $t_{0}$ is given by
\begin{equation}
   \label{eqn:mass}
   M(R_{c_{1}}, t_{0}) \, = \, \beta\gamma(G\mu)^{2}z_{eq}^{-1/2}\Big(\frac{t_{0}}{G}\Big) \, .
\end{equation}
To obtain a feeling for the meaning of this expression, let us insert the 
values of $t_{0}$ and $G$. We then have
\begin{equation}
   \label{eqn:t0/G}
   \frac{t_{0}}{G} \, \sim \, 10^{23}M_{\odot} \, ,
\end{equation}
where $M_{\odot}$ stands for the solar mass.

In our analysis, we take the peak mass $M_{c}$ (the mass where
the observed globular cluster mass function for our Milky Way galaxy peaks) 
to fix our only free parameter, the string tension. The mass function scales 
as $M^{-5/2}$ for $M > M_{c}$ (which follows directly from the string loop 
distribution (\ref{eqn:numberdensity})). We predict a linear decay for 
$M < M_{c}$. This comes about since the loop radius distribution is constant and 
loops with radius smaller than $R_{c_{1}}$ live only a fraction of a 
Hubble time step which scales linearly with $R$.

In Fig. \ref{fig:histogram1_2}, a comparison of the predicted mass function 
(the solid lines) with the observed distribution of globular clusters in the 
Milky Way (histogram values) complied from \cite{GlobularClusterParameters} is 
made. To obtain the theoretical curve, we take the comoving number density 
$n(R,t_{eq})z_{eq}^{3}$ of loops (where $n(R,t)$ is given in 
(\ref{eqn:numberdensity})), multiply the result by the concentration factor $F$, 
allow each loop to grow in mass by a factor of $z_{eq}$ (independent spherical 
accretion), and convert this into a mass distribution $n(M,t_{0})$, while taking 
into account the Jacobian of the transformation from $R$ to $M$. The result is 
then multiplied by the bin size $\delta M=fM$, where $f$ is a number, and by 
the volume $V$ of the Milky Way galaxy. We obtain the following peak number 
density bin using (\ref{eqn:mass})
\begin{equation}
   \label{peaknumberdensitybin}
   \delta N \, = \, 
   NFf\alpha^{5/2}\beta^{-5/2}\gamma^{-3/2}(G\mu)^{-3/2}z_{eq}^{3/2}t_{0}^{-3}V \, ,
\end{equation}
where $z(t)$ is the cosmological redshift at time $t$.

All of the calculations summarized in the this section assumed that 
cosmic strings loops are created and remain at rest. In the
following two sections, we will study the effect of velocities (translational 
center of mass motion) through two different analyses. 

In our first analysis (discussed in the following section), 
we will compute the mass function of objects
which accrete onto loops with velocities low enough such that the
loop center moves a smaller physical distance than the  physical radius 
of the shell which would be collapsing onto the loop if it had been
stationary. If the loop moves further than this, we assume that no globular 
clusters forms.

In our second analysis (to be discussed in the next to following
section) we study the accretion of matter onto moving
loops and keep only objects which are sufficiently spherical.
We find that both conditions give similar resulting mass functions for
globular clusters, the one from the second condition being slightly
higher.

In the following analyses, we are neglecting the rocket effect which is 
the effect of anisotropy in the loop gravitational radiation that causes 
loops to accelerate as they decay. We show in section \ref{sec:5}
that this is indeed a good approximation.

\section{\label{sec:3}Effect of Cosmic String Velocity: First Analysis}

Accretion onto a cosmic string loop can be studied using the
Zel'dovich approximation. As shown in e.g. in \cite{Pagano}, 
the physical distance $h(q,t_{0})$ from the center of a string loop to the mass shell
which is ``turning around'' (i.e. becoming gravitationally
bound) at the present time $t_{0}$  is given by
\begin{equation}
   \label{eqn:h}
   h(R,t_{0}) \, = \, 
(\tfrac{9}{5})^{1/3}\beta^{1/3}(G\mu)^{1/3}z_{eq}^{1/3}t_{0}^{2/3}R^{1/3} \, .
\end{equation}
On the other hand, for an initial physical velocity $v_{i}$, the distance a 
loop of radius $R$ has moved by the present time is
\begin{align}
   \label{eqn:deltar}
   \Delta r(R)&=a(t_{0})\int_{t_{eq}}^{t_{0}}\Big(\frac{a(t_{i})}{a(t)}\Big)^{2}\frac{v_{i}}{a(t_{i})} dt \nonumber \\
   &=3\alpha^{-1/2}\beta^{1/2}z_{eq}^{1/4}t_{0}^{1/2}R^{1/2}v_{i} \, .
\end{align}

\begin{figure}[ht]
\centering
\includegraphics[width=0.48\textwidth]{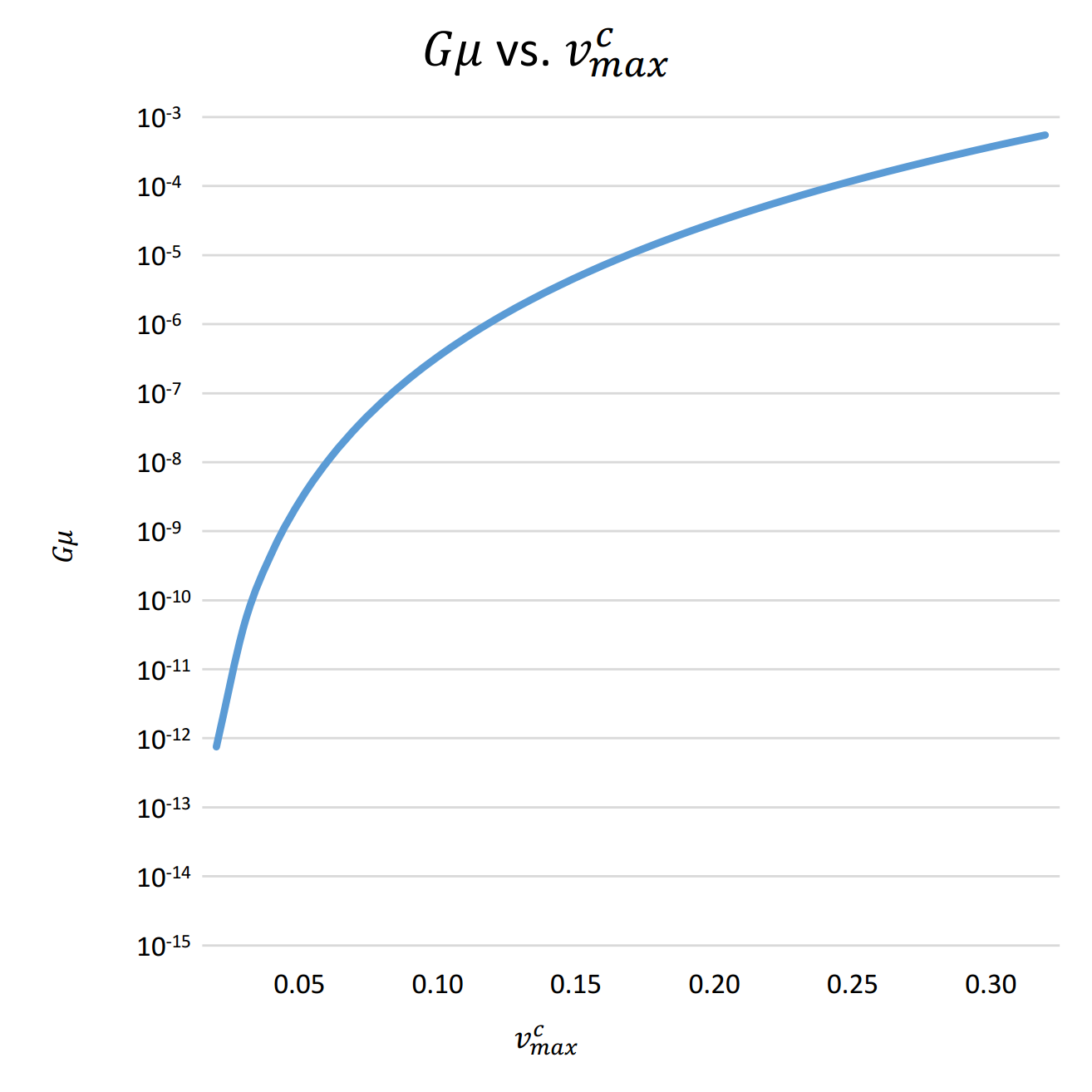}
\caption{Analysis 1 - Relation between $G\mu$ and $v_{\text{max}}^{c}$ (\ref{eqn:vmaxc_1}). 
The horizontal axis is velocity on a linear scale and the vertical axis gives the $G\mu$ on a 
logarithmic scale.}
\label{fig:graph2_1_3}
\end{figure}

In the present analysis we will only count the number of string loops for which
\be
 \Delta r(R) \, < \, h(R,t_{0}) \, ,
 \ee
 and we assume that the accretion onto faster moving loops is not effective at
 producing compact globular clusters. Making use of (\ref{eqn:h}) and
 (\ref{eqn:deltar}) we obtain an upper bound on the initial velocity:
\begin{equation}
   \label{eqn:vi_1}
   v_{i} \, < \, 
   (\tfrac{1}{15})^{1/3}\alpha^{1/2}\beta^{-1/6}(G\mu)^{1/3}z_{eq}^{1/12}t_{0}^{1/6}R^{-1/6} \, .
\end{equation}
Taking the distribution of initial velocities in each of three spatial directions to be a 
step function of width $v_{\text{max}}$ leads to the following probability that a string
loop will satisfy the condition (\ref{eqn:vi_1}):
\begin{equation}
   \label{eqn:P}
   \mathcal{P}(v) \, = \, \tfrac{1}{3}v_{i}^{3}v_{\text{max}}^{-3} \, ,
\end{equation}
where $v_{\text{max}}$ is the maximum velocity of cosmic strings determined through 
cosmic string evolution simulations. Taking the integral of the velocity distribution from 
zero to the upper bound on $v_{i}$, the rate of globular cluster formation becomes 
suppressed by a multiplicative factor $\mathcal{S}(R)$:
\begin{equation}
   \label{eqn:factor_1}
   \mathcal{S}(R) \, = \, 
   \tfrac{1}{45}\alpha^{3/2}\beta^{-1/2}(G\mu)z_{eq}^{1/4}t_{0}^{1/2}v_{\text{max}}^{-3}R^{-1/2} \, .
\end{equation}
The distribution has a cutoff when $\mathcal{S}(R)=1$ leading to the critical radius $R_{c_{2}}$:
\begin{equation}
   \label{eqn:Rc2_1}
   R_{c_{2}} \, = \,
   (\tfrac{1}{45})^{2}\alpha^{3}\beta^{-1}(G\mu)^{2}z_{eq}^{1/2}t_{0}v_{\text{max}}^{-6} \, .
\end{equation}
Hence for values of $R < R_{c_{2}}$, there is no suppression from velocity effects and 
for $R > R_{c_{2}}$ the suppression is given above by (\ref{eqn:factor_1}). Setting 
$R_{c_{2}} = R_{c_{1}}$ we obtain a critical $v_{\text{max}}^{c}$ given by
\begin{equation}
   \label{eqn:vmaxc_1}
   v_{\text{max}}^{c} \, = \, 
   (\tfrac{1}{45})^{1/3}\alpha^{1/2}\beta^{-1/6}\gamma^{-1/6}(G\mu)^{1/6}z_{eq}^{1/3} \, .
\end{equation}
Thus, for $v_{\text{max}} < v_{\text{max}}^{c}$ we have $R_{c_{2}} > R_{c_{1}}$ and
hence the mass function of predicted globular clusters from string loops will not
change near the peak position compared to what was obtained in \cite{us} neglecting the presence
of string loop velocities. On the other hand, if $v_{\text{max}} > v_{\text{max}}^{c}$
then the mass function will be suppressed near the peak position.
The relation of $v_{\text{max}}^{c}$ and $G\mu$ is illustrated in Fig. \ref{fig:graph2_1_3}. 

Taking into account the suppression factor when performing the calculations outlined in 
section \ref{sec:2}, we obtain the following histogram of predicted number of globular
clusters:
\begin{align}
   \label{eqn:numberdensitybin1_1}
   &\underline{\text{For}\;R_{c_{2}}>R_{c_{1}}:}\\
   &\begin{aligned}
      \quad\delta N=&NFf\alpha^{5/2}\beta^{-5/2}\gamma^{-3/2}\\
      &\quad\times(G\mu)^{-3/2}z_{eq}^{3/2}t_{0}^{-3}V& &\text{at}\;R_{c_{1}}\\
      \quad\delta N=&(\tfrac{1}{45})^{-3}NFf\alpha^{-2}\beta^{-1} \\
      &\quad\times(G\mu)^{-3}z_{eq}^{-3/2}t_{0}^{-3}v_{\text{max}}^{9}V& &\text{at}\;R_{c_{2}}
   \end{aligned} \nonumber \\
   \label{eqn:numberdensitybin2_1}
   &\underline{\text{For}\;R_{c_{2}}<R_{c_{1}}:}\\
   &\begin{aligned}
      \quad\delta N=&\tfrac{1}{45}NFf\alpha^{4}\beta^{-3}\gamma^{-2}\\
      &\quad\times(G\mu)^{-1}z_{eq}^{5/2}t_{0}^{-3}v_{\text{max}}^{-3}V & &\text{at}\;R_{c_{1}}\\
      \quad\delta N=&(\tfrac{1}{45})^{2}NFf\alpha^{11/2}\beta^{-7/2}\gamma^{-5/2}\\
      &\quad\times(G\mu)^{-1/2}z_{eq}^{7/2}t_{0}^{-3}v_{\text{max}}^{-6}V& &\text{at}\;R_{c_{2}}\\
   \end{aligned} \nonumber
\end{align}

Notice that for $R_{c_{1}} < R_{c_{2}}$, the mass scales as $M^{-3}$ for $R > R_{c_{2}}$. 
In the radius $R$ interval between $R_{c_{1}}$ and $R_{c_{2}}$ the mass function
scales as $M^{-5/2}$ as it does in the absence of velocity effects, and for masses smaller 
than $M_{c}$, a linear decay is predicted by the same reasoning as in section \ref{sec:2}. 
For $R_{c_{1}} > R_{c_{2}}$, the mass function scales as $M^{-3}$ for $R > R_{c_{1}}$, 
as $M^{-1/2}$ for $R_{c_{2}} < R< R_{c_{1}}$ and decays linear for $R < R_{c_{2}}$.

\begin{figure*}
\centering
\includegraphics[width=\textwidth]{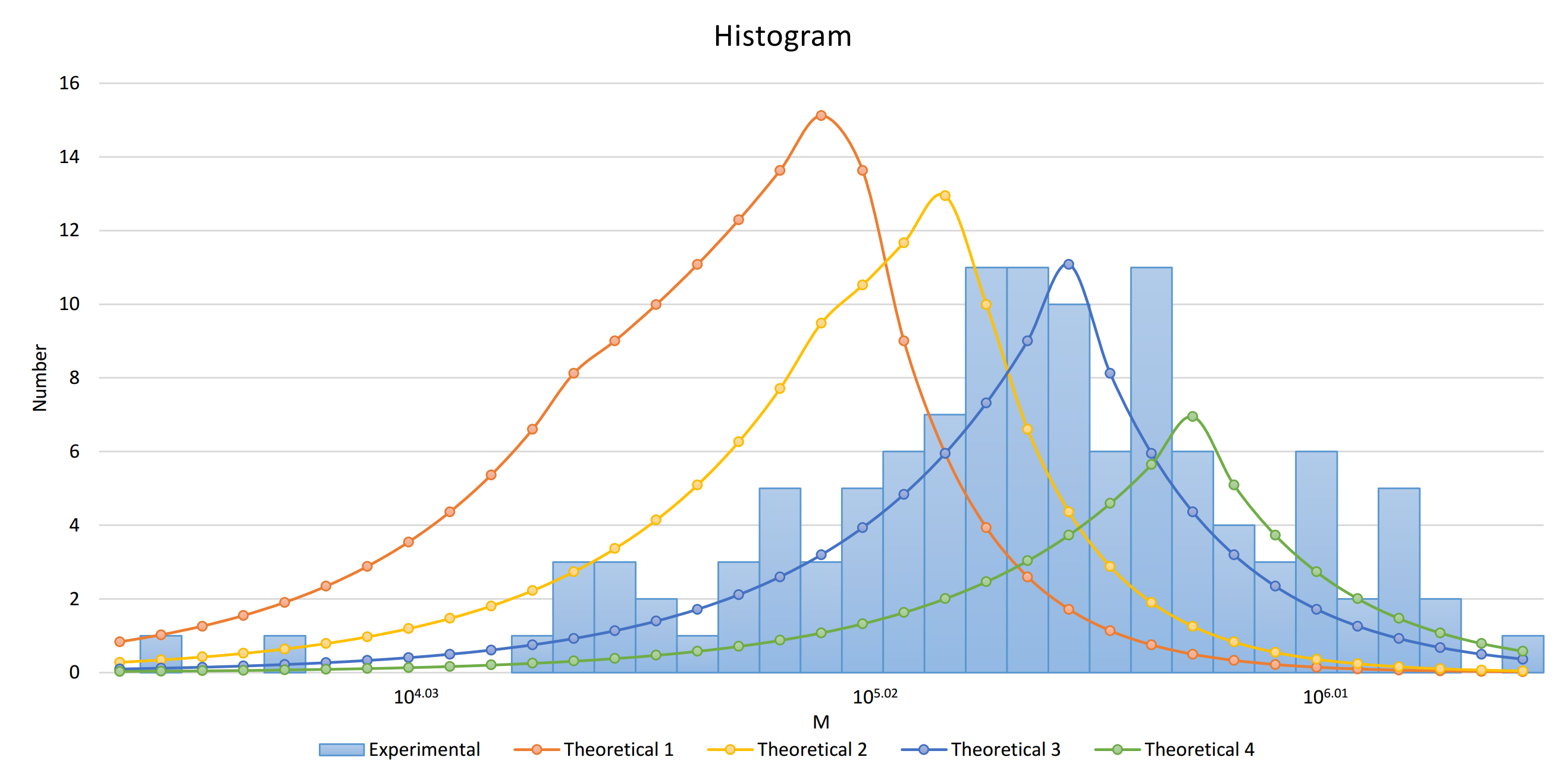}
\caption{Analysis 1 - Dependence of the mass function on $G\mu$ at 
$v_{\text{max}}=3.00\times10^{-2}$. The horizontal axis is mass on a logarithmic scale, 
the vertical axis gives the number density on a linear scale. The histogram shows data 
taken from \cite{GlobularClusterParameters}. The curves shown are for 
$G\mu=2.92\times10^{-10}$, $G\mu=3.98\times10^{-10}$, $G\mu=5.43\times10^{-10}$, 
and $G\mu=7.41\times10^{-10}$ (in increasing order of mass at the peak position). 
Notice that for the red and yellow solid curves, $R_{c_{2}}<R_{c_{1}}$, for the blue 
solid curve $R_{c_{2}}=R_{c_{1}}$ and for the green solid curve $R_{c_{2}}>R_{c_{1}}$. 
The blue solid curve minimizes $\chi^{2}$ for this particular $v_{\text{max}}$.}
\label{fig:histogram2_1_1}
\end{figure*}

\begin{figure*}
\centering
\includegraphics[width=\textwidth]{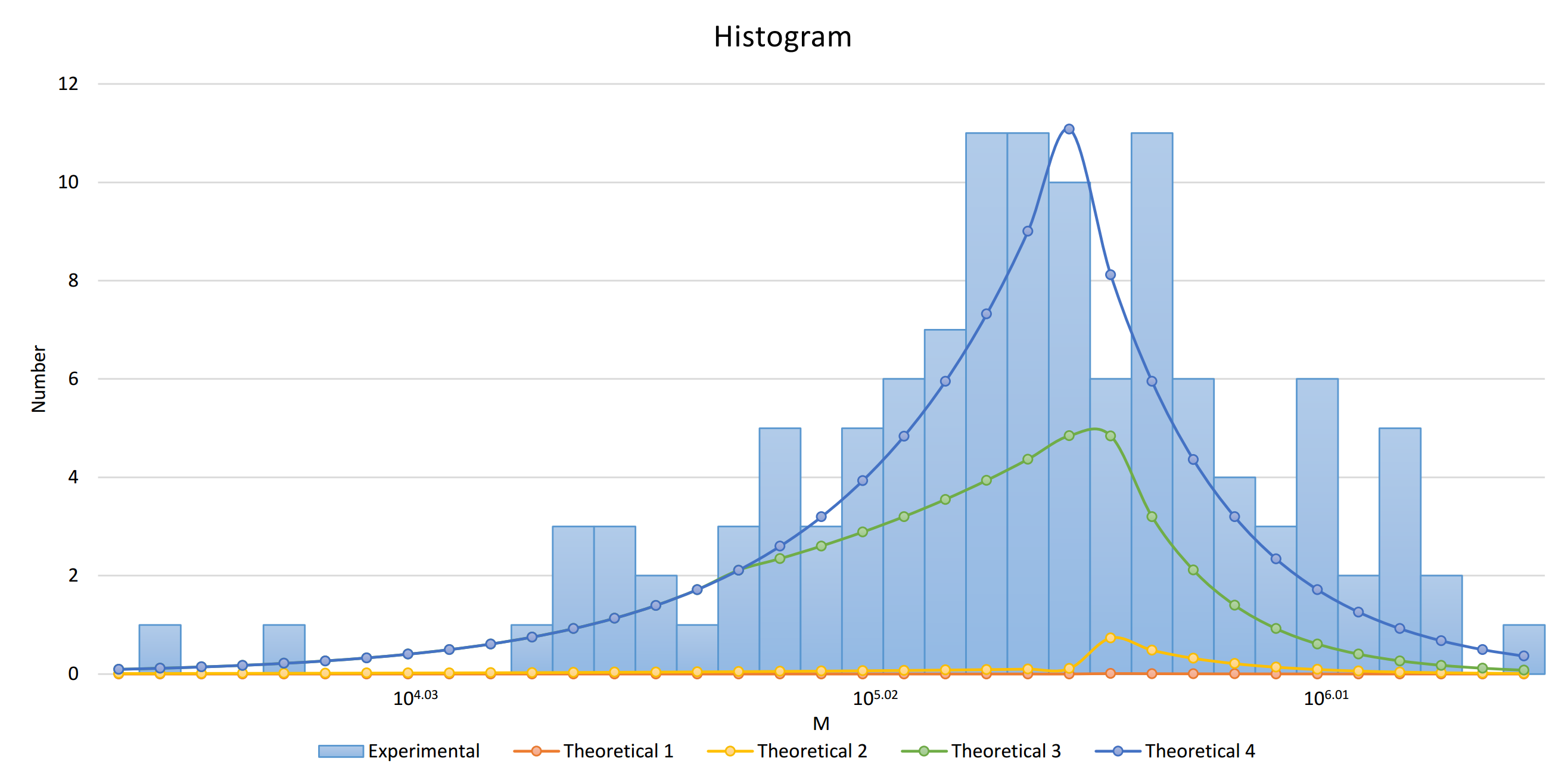}
\caption{Analysis 1 - Dependence of the mass function at $G\mu=5.43\times10^{-10}$ on 
$v_{\text{max}}$. The curves are shown for $v_{\text{max}}=1.00\times10^{-1}$ 
(Theoretical 1), $v_{\text{max}}=6.46\times10^{-2}$ (Theoretical 2), 
$v_{\text{max}}=3.44\times10^{-2}$ (Theoretical 3) and $v_{\text{max}}=3.00\times10^{-2}$ 
(Theoretical 4). Notice that for all $v_{\text{max}}<3.00\times10^{-2}$ we would obtain the 
blue curve. The axes and data are the same as in the previous figure.}
\label{fig:histogram2_1_2}
\end{figure*}

In Fig. \ref{fig:histogram2_1_1}, we show that varying $G\mu$ shifts the peak position and amplitude of the mass for a fixed $v_{\text{max}}^{c}$. For $R_{c_{2}}>R_{c_{1}}$, initial velocity effects are negligible.  However, for  $R_{c_{2}}<R_{c_{1}}$ but very close to the value of $R_{c_{1}}$ there is a slight suppression in the region $ R_{c_{2}}<R< R_{c_{1}}$ from velocity effects.\\
\indent In Fig. \ref{fig:histogram2_1_2}, we consider $G\mu=5.43\times10^{-10}$ which minimizes $\chi^{2}$ in Fig. \ref{fig:histogram1_2}, we find from varying $v_\text{max}$ that for $v_\text{max}<3.00\times10^{-2}$, velocity has little effect on mass distribution of globular clusters in the Milky Way galaxy. However, for $v_\text{max}\gg3.00\times10^{-2}$ we will not obtain a mass distribution.

\section{\label{sec:4}Effect of Cosmic String Velocity: Second Analysis}

In the previous section we estimated the range of velocities for which spherical accretion onto
a loop is a good approximation. On the other hand, accretion onto a moving loop can
also be studied by means of the Zel'dovich approximation. This analysis is technically
a bit more complicated than in the case of spherical accretion, but the study has been
carried out in \cite{Accretion}.  The result is that half of the turnaround mass from a 
string with some initial velocity is within a region which can be approximated by a paraboloid 
of radius $r=b^{1/3}d_{i}$ and height $h=4b^{1/3}d_{i}$),
where (for loops born before the time of equal matter and radiation)
\be \label{bparameter}
b(t) \, = \, \frac{1}{15} \frac{G m}{v_{eq}^3 t_{eq}} a(t) \, ,
\ee
where $m$ is the mass of the loop.
Note that since the accretion effectively starts at $t_{eq}$, it is the loop velocity $v_{eq}$ at
that time which enters the formula.

The value of the mass enclosed in this region is:
\begin{equation}
   \label{eqn:Mns_half}
   M_{\text{ta}_{1/2}}^{\text{ns}}(t)=\tfrac{3}{5}ma(t)\qquad\text{for}\;b(t)\ll1 \, .
\end{equation}
Assuming that the accreted mass has uniform density, we find that density $\rho$ is given by
\begin{equation}
   \label{eqn:density}
   \rho=\frac{M}{V}=\tfrac{3}{10}\pi^{-1}b^{-1}d_{i}^{-3}ma(t)\qquad\text{for}\;b(t)\ll1 \, .
\end{equation}
Approximating the other half of the accreted mass to be spherical, the total mass is given by
\begin{equation}
   \label{eqn:Mns_total}
   M_{\text{t}}^{\text{ns}}=\tfrac{4}{5}ma(t) \, .
\end{equation}
Comparing this to the mass from spherical accretion:
\begin{equation}
   \label{eqn:Ms_total}
   M_{\text{t}}^{\text{s}}=\tfrac{2}{5}ma(t)
\end{equation}
we see non-spherical accretion results in a mass that is larger by a factor of two.

\begin{figure}[ht]
\centering
\includegraphics[width=0.48\textwidth]{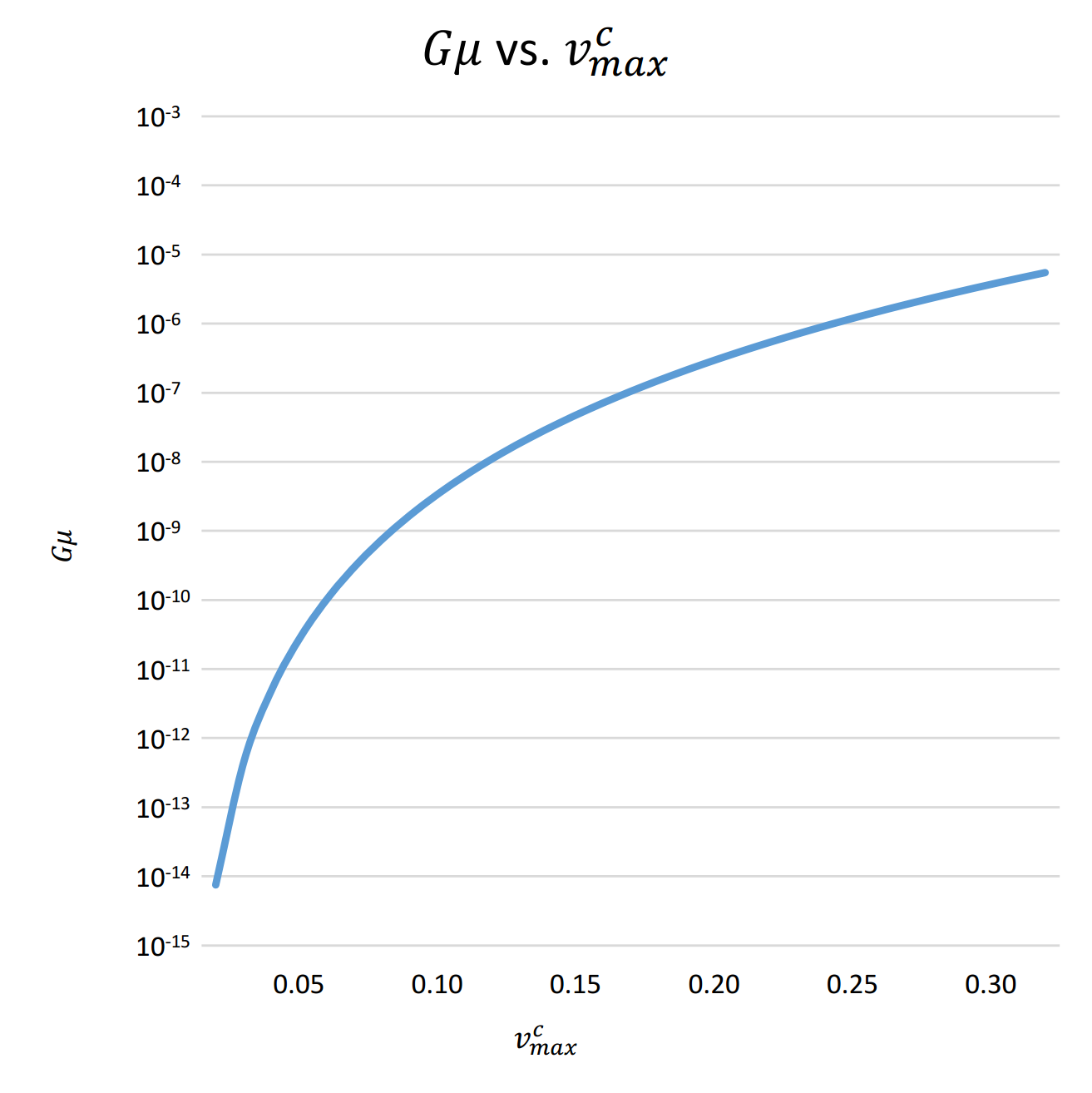}
\caption{Analysis 2 - Relation between $G\mu$ and $v_{\text{max}}^{c}$ (\ref{eqn:vmaxc_2}). 
The horizontal axis is velocity on a linear scale and the vertical axis gives the $G\mu$ on a 
logarithmic scale.}
\label{fig:graph2_2_3}
\end{figure}

Accretion is roughly spherical when the loop accretion sphericity parameter $b(t)$ 
as defined in (\ref{bparameter}) is larger than one. In this analysis, we will consider a 
slightly lower bound by setting $b(t) > 10^{-1}$. Using this condition, we obtain an 
equation for $v_{eq}$:
\begin{equation}
   \label{veq}
   v_{eq}<(\tfrac{2}{3})^{1/3}\beta^{1/3}(G\mu)^{1/3}z_{eq}^{5/6}t_{0}^{-1/3}R^{1/3} \, .
\end{equation}
Red-shifting the velocity to the time of loop formation in the radiation dominated era we obtain:
\begin{equation}
   \label{eqn:vi_2}
   v_{i}<(\tfrac{2}{3})^{1/3}\alpha^{1/2}\beta^{-1/6}(G\mu)^{1/3}z_{eq}^{1/12}t_{0}^{1/6}R^{-1/6} \, .
\end{equation}
Notice that the upper bound on the initial velocity in this analysis differs by only factor of 
$10^{1/3}$ from the upper bound found in the first analysis. From here, performing the 
same steps as in Analysis 1 would obtain results that are larger by a factor $10^{1/3}$. 
This can be seen clearly by determining the new $v_{\text{max}}^{c}$:
\begin{equation}
   \label{eqn:vmaxc_2}
   v_{\text{max}}^{c}=(\tfrac{2}{9})^{1/3}\alpha^{1/2}\beta^{-1/6}\gamma^{-1/6}(G\mu)^{1/6}z_{eq}^{1/3} \, .
\end{equation}
The numerical results are presented in Fig. (\ref{fig:graph2_2_3}).

\section{\label{sec:5}The Rocket Effect}

The {\it rocket effect} is an effect which causes the string loop to
accelerate as a consequence of anisotropic gravitational radiation.
Let us assume that the string loop has a center of mass velocity
$v_{eq}$ at $t_{eq}$. Hubble expansion will lead to a decrease
in the velocity whereas the anisotropic gravitational radiation will
tend to cause an increase. Which effect wins depends on time.

The equation of motion for the loop velocity taking into account
Hubble expansion and anisotropic gravitational radiation is (see e.g.
Appendix A of \cite{Loeb} - we have generalized the result to times
before $t_{eq}$)
\begin{equation} \label{rocket}
   v(t) \, = \, v_i \frac{z(t)}{z(t_i)} 
   + \left(\frac{3\Gamma_P(G\mu)^2M_i}{5m}\right)\left(\frac{t}{t_i}-\frac{t_i^{1/2}}{t^{1/2}}\right) \, ,
\end{equation}
where $\Gamma_{P}$ is the coefficient of anisotropic gravitational radiation, $M_i$ is 
the mass at the loop formation time $t_i$ and $m$ is the mass at a later time.
We will take $\Gamma_{P}\simeq10$.  This value was estimated by 
numerical simulations using the Kibble-Turok loop solutions \cite{intersection}. 

We now ask the question whether the rocket effect can have an important effect
on gravitational accretion by the loops which dominate the globular cluster
mass function. Thus, we are interested in loops with initial radius $R = \gamma G \mu t_{eq}$.
If the second term on the right-hand side of (\ref{rocket}) is smaller than the first
term at the time $t = t_{eq}$, then we argue that the rocket effect will be negligible
since even if at very late times the string loop starts to accelerate away from the
initial position of the loop, the nonlinear object seeded by the loop is already
in place one Hubble time after $t_{eq}$. Making use of (\ref{rocket}) we then
obtain the condition
\be
v_i (G \mu)^{-{1/2}} \, > \frac{3}{5} \Gamma_P \alpha^{3/2} \beta^{-3/2} \gamma^{-3/2} \, ,
\ee
which, if satisfied, guarantees that the rocket effect is negligible. But for the values
of $G \mu$ smaller than the current observational bound this condition is obviously
satisfied. Hence, we conclude that the rocket effect does not effect our
analysis.

\section{\label{sec:6}Conclusions and Discussion}

In this paper we have considered the effects of string loop velocities 
on the model for the explanation of the origin of globular clusters proposed in \cite{us}
in which it is proposed that globular clusters are seeded by cosmic string loops. 

We see that velocities play a small, but noticeable role in the accretion of matter during 
the matter-dominated era. Cosmic string loops, born in the radiation era at $t_{i}$ with 
initial velocity $v_{i}$ will travel a certain given by (\ref{eqn:deltar}). Demanding
that this distance be smaller than the total size of the structure which accretes
around a static loop, denoted in this paper by $h(R,t_0)$, leads to an upper bound
on the string velocities. In our Analysis 1 we assumed that loops with larger
velocities do not give rise to globular clusters. We computed the corrections
to the mass function obtained using this hypothesis, as a function of
the maximal velocity of string loops. Our theory predicts a cutoff velocity. 
For velocities smaller than $v_{\text{max}}^{c}$, velocity effects are negligible.

An improved analysis (Analysis 2) can be obtained by taking into account the
non-spherical accretion of matter about a moving string loop, and demanding that
the resulting nonlinear object is sufficiently spherical. The results of this analysis
were very similar to those obtained using Analysis 1.

Finally, we have shown that the {\it rocket effect} does not affect our
analysis.

Aside from globular clusters, it is noteworthy to mention that analyses discussed in this 
report may also be applied to ultra-compact-mini-halos, as they too can be manifestations 
of cosmic string loop accretion \cite{UCMH}. Work on this topic is in progress.

\acknowledgements{We wish to thank K. Olum and A. Vilenkin for useful
discussions and encouragement. RB is supported by an NSERC Discovery Grant, and by funds 
from the Canada Research Chair program.}


\begin{thebibliography}{99}

\bibitem{us}
A.~Barton, R.~H.~Brandenberger and L.~Lin,
  ``Cosmic Strings and the Origin of Globular Clusters,''
  JCAP {\bf 1506}, no. 06, 022 (2015)
  [arXiv:1502.07301 [astro-ph.CO]].

\bibitem{Kibble}
T.~W.~B.~Kibble,
  ``Phase Transitions In The Early Universe,''
  Acta Phys.\ Polon.\  B {\bf 13}, 723 (1982);\\
  T.~W.~B.~Kibble,
  ``Some Implications Of A Cosmological Phase Transition,''
  Phys.\ Rept.\  {\bf 67}, 183 (1980).

\bibitem{CSearly}
N.~Turok and R.~H.~Brandenberger,
  ``Cosmic Strings And The Formation Of Galaxies And Clusters Of Galaxies,''
  Phys.\ Rev.\ D {\bf 33}, 2175 (1986);\\
H. Sato, ``Galaxy Formation by Cosmic Strings,''
  Prog. Theor. Phys.\  {\bf 75}, 1342 (1986);\\
A. Stebbins, ``Cosmic Strings and Cold Matter'',
  Ap. J. (Lett.) {\bf 303}, L21 (1986).  
  
\bibitem{CSlimit}
C.~Dvorkin, M.~Wyman and W.~Hu,
  ``Cosmic String constraints from WMAP and the South Pole Telescope,''
  Phys.\ Rev.\ D {\bf 84}, 123519 (2011)
  [arXiv:1109.4947 [astro-ph.CO]];\\
  P.~A.~R.~Ade {\it et al.}  [Planck Collaboration],
  ``Planck 2013 results. XXV. Searches for cosmic strings and other topological defects,''
  Astron.\ Astrophys.\  {\bf 571}, A25 (2014)
  [arXiv:1303.5085 [astro-ph.CO]].

 \bibitem{others}
L.~Pogosian, S.~H.~H.~Tye, I.~Wasserman and M.~Wyman,
  ``Observational constraints on cosmic string production during brane
  inflation,''
  Phys.\ Rev.\  D {\bf 68}, 023506 (2003)
  [Erratum-ibid.\  D {\bf 73}, 089904 (2006)]
  [arXiv:hep-th/0304188];\\
M.~Wyman, L.~Pogosian and I.~Wasserman,
  ``Bounds on cosmic strings from WMAP and SDSS,''
  Phys.\ Rev.\  D {\bf 72}, 023513 (2005)
  [Erratum-ibid.\  D {\bf 73}, 089905 (2006)]
  [arXiv:astro-ph/0503364];\\
A.~A.~Fraisse,
  ``Limits on Defects Formation and Hybrid Inflationary Models with
  Three-Year WMAP Observations,''
  JCAP {\bf 0703}, 008 (2007)
  [arXiv:astro-ph/0603589];\\
U.~Seljak, A.~Slosar and P.~McDonald,
  ``Cosmological parameters from combining the Lyman-alpha forest with CMB,
  galaxy clustering and SN constraints,''
  JCAP {\bf 0610}, 014 (2006)
  [arXiv:astro-ph/0604335];\\
  R.~A.~Battye, B.~Garbrecht and A.~Moss,
  ``Constraints on supersymmetric models of hybrid inflation,''
  JCAP {\bf 0609}, 007 (2006)
  [arXiv:astro-ph/0607339];\\
R.~A.~Battye, B.~Garbrecht, A.~Moss and H.~Stoica,
  ``Constraints on Brane Inflation and Cosmic Strings,''
  JCAP {\bf 0801}, 020 (2008)
  [arXiv:0710.1541 [astro-ph]];\\
N.~Bevis, M.~Hindmarsh, M.~Kunz and J.~Urrestilla,
  ``CMB power spectrum contribution from cosmic strings using  field-evolution
  simulations of the Abelian Higgs model,''
  Phys.\ Rev.\  D {\bf 75}, 065015 (2007)
  [arXiv:astro-ph/0605018];\\
N.~Bevis, M.~Hindmarsh, M.~Kunz and J.~Urrestilla,
  ``Fitting CMB data with cosmic strings and inflation,''
  Phys.\ Rev.\ Lett.\  {\bf 100}, 021301 (2008)
  [astro-ph/0702223 [ASTRO-PH]];\\
R.~Battye and A.~Moss,
  ``Updated constraints on the cosmic string tension,''
 Phys.\ Rev.\ D {\bf 82}, 023521 (2010)
  [arXiv:1005.0479 [astro-ph.CO]].
  
 \bibitem{VSimulations}
 A.~Albrecht and N.~Turok,
  ``Evolution Of Cosmic Strings,''
  Phys.\ Rev.\ Lett.\  {\bf 54}, 1868 (1985);\\
D.~P.~Bennett and F.~R.~Bouchet,
  ``Evidence For A Scaling Solution In Cosmic String Evolution,''
  Phys.\ Rev.\ Lett.\  {\bf 60}, 257 (1988);\\
B.~Allen and E.~P.~S.~Shellard,
  ``Cosmic String Evolution: A Numerical Simulation,''
  Phys.\ Rev.\ Lett.\  {\bf 64}, 119 (1990);\\
C.~Ringeval, M.~Sakellariadou and F.~Bouchet,
  ``Cosmological evolution of cosmic string loops,''
  JCAP {\bf 0702}, 023 (2007)
  [arXiv:astro-ph/0511646];\\
V.~Vanchurin, K.~D.~Olum and A.~Vilenkin,
  ``Scaling of cosmic string loops,''
  Phys.\ Rev.\  D {\bf 74}, 063527 (2006)
  [arXiv:gr-qc/0511159];\\
  L.~Lorenz, C.~Ringeval and M.~Sakellariadou,
  ``Cosmic string loop distribution on all length scales and at any redshift,''
  JCAP {\bf 1010}, 003 (2010)
  [arXiv:1006.0931 [astro-ph.CO]];\\
J.~J.~Blanco-Pillado, K.~D.~Olum and B.~Shlaer,
  ``Large parallel cosmic string simulations: New results on loop production,''
  Phys.\ Rev.\ D {\bf 83}, 083514 (2011)
  [arXiv:1101.5173 [astro-ph.CO]];\\
  J.~J.~Blanco-Pillado, K.~D.~Olum and B.~Shlaer,
  ``The number of cosmic string loops,''
  Phys.\ Rev.\ D {\bf 89}, no. 2, 023512 (2014)
  [arXiv:1309.6637 [astro-ph.CO]].
  
 \bibitem{Onescale1}
 A.~Vilenkin,
  ``Cosmic Strings,''
  Phys.\ Rev.\ D {\bf 24}, 2082 (1981).

 \bibitem{Onescale2}
 T.~W.~B.~Kibble,
  ``Evolution of a system of cosmic strings,''
  Nucl.\ Phys.\ B {\bf 252}, 227 (1985)
  [Nucl.\ Phys.\ B {\bf 261}, 750 (1985)].
 
\bibitem{grav}
  T.~Vachaspati and A.~Vilenkin,
  ``Gravitational Radiation from Cosmic Strings,''
  Phys.\ Rev.\ D {\bf 31}, 3052 (1985).
  
 \bibitem{CSrevs}
A. Vilenkin and E.P.S. Shellard, \textit{Cosmic Strings and other
Topological Defects} (Cambridge Univ. Press, Cambridge, 1994);\\
M.~B.~Hindmarsh and T.~W.~B.~Kibble,
  ``Cosmic strings,''
  Rept.\ Prog.\ Phys.\  {\bf 58}, 477 (1995)
  [arXiv:hep-ph/9411342];\\
R.~H.~Brandenberger,
  ``Topological defects and structure formation,''
  Int.\ J.\ Mod.\ Phys.\ A {\bf 9}, 2117 (1994)
  [arXiv:astro-ph/9310041].
  
\bibitem{Zeldovich}
Y.~.B.~Zeldovich,
  ``Gravitational instability: An Approximate theory for large density perturbations,''
  Astron.\ Astrophys.\  {\bf 5}, 84 (1970).

  \bibitem{GlobularClusterParameters}
O.~Y.~Gnedin and J.~P.~Ostriker,
  ``Destruction of the galactic globular cluster system,''
  Astrophys.\ J.\  {\bf 474}, 223 (1997)
  [astro-ph/9603042].

\bibitem{Pagano} 
  M.~Pagano and R.~Brandenberger,
  ``The 21cm Signature of a Cosmic String Loop,''
  JCAP {\bf 1205}, 014 (2012)
  [arXiv:1201.5695 [astro-ph.CO]].

\bibitem{Accretion}   
E. Bertschinger,
    ``Cosmological Accretion Wakes'',
    Astrophys. J. {\bf 316}, 489 (1987).

\bibitem{Loeb}
B.~Shlaer, A.~Vilenkin and A.~Loeb,
  ``Early structure formation from cosmic string loops,''
  JCAP {\bf 1205}, 026 (2012)
  [arXiv:1202.1346 [astro-ph.CO]].
  
\bibitem{intersection}
T.~W.~B.~Kibble and N.~Turok,
  ``Selfintersection of Cosmic Strings,''
  Phys.\ Lett.\ B {\bf 116}, 141 (1982).
  
\bibitem{UCMH}
M.~Anthonisen, R.~Brandenberger and P.~Scott,
  ``Constraints on cosmic strings from ultracompact minihalos,''
  Phys.\ Rev.\ D {\bf 92}, no. 2, 023521 (2015)
  [arXiv:1504.01410 [astro-ph.CO]].
  
\end{thebibliography}
\end{document}